\documentclass[11pt,a4paper,twoside,fleqn]{article}
\usepackage{color,graphicx,natbib,lscape,here,geometry,setspace,multirow,enumitem}
\usepackage[dvipsnames]{xcolor}
\usepackage{graphicx}
\usepackage{authblk}
\usepackage{silence}
\usepackage{rotating}
\usepackage{multirow}

\usepackage{booktabs}
\usepackage{diagbox}
\usepackage{color}
\usepackage{setspace}
\usepackage{algorithm2e}
\usepackage{enumitem}
\usepackage{tikz}

\definecolor{nblue}{HTML}{000660}
\usepackage[colorlinks=true,urlcolor=nblue,linkcolor=nblue,citecolor=nblue]{hyperref}
\usepackage{bigstrut}

\usepackage{tabularx}
\usepackage{threeparttable}
\usepackage{dcolumn}
\newcolumntype{d}[1]{D{.}{.}{#1}}

\usepackage{array}
\newcolumntype{C}[1]{>{\centering\arraybackslash}p{#1}}

\usepackage{etoolbox} 
\makeatletter
\patchcmd{\BR@backref}{\newblock}{\newblock[}{}{}
\patchcmd{\BR@backref}{\par}{]\par}{}{}
\makeatother

\usepackage{pdflscape}
\usepackage{afterpage}

\usepackage{longtable}
\usepackage{array}

\usepackage{comment}
\usepackage{mathtools}

\usepackage[hang,flushmargin]{footmisc} 

\usepackage[hang]{footmisc}
\usepackage[british]{babel}

\pdfoutput=1
\usepackage{float}
\restylefloat{table}

\usepackage[title,titletoc]{appendix}
\makeatletter

\renewenvironment{appendices}{%
    \begin{oldappendices}%
    \renewcommand{\thefigure}{\ifnum \c@section>\z@ \thesection.\fi\@arabic\c@figure}%
    \@addtoreset{figure}{section}%
    \renewcommand{\thetable}{\ifnum \c@section>\z@ \thesection.\fi\@arabic\c@table}%
    \@addtoreset{table}{section}}{%
    \end{oldappendices}%
}\makeatother

\usepackage{titlesec} 
\titleformat{\section}[block]{\large}{\thesection. }{0em}{\MakeUppercase} 
\titleformat{\subsection}[block]{\large}{\thesubsection. }{0em}{\itshape} 
\titleformat{\subsubsection}[block]{\large}{}{0em}{\itshape} 

\let\natbibcitet\citet
\renewcommand\citet{\bibpunct{(}{)}{,}{a}{,}{,}\natbibcitet}
\let\natbibcitep\citep
\renewcommand\citep{\bibpunct{(}{)}{;}{a}{,}{;}\natbibcitep}
\newcommand{\bi}{\begin{itemize}}
\newcommand{\ei}{\end{itemize}}
\newcommand{\be}{\begin{equation}}
\newcommand{\ee}{\end{equation}}
\defcitealias{ieo14}{IEO, 2014}

\long\def\symbolfootnote[#1]#2{\begingroup%
\def\thefootnote{\fnsymbol{footnote}}\footnote[#1]{#2}\endgroup}

\makeatletter
\def\ubar#1{\underline{\sbox\tw@{$#1$}\dp\tw@\z@\box\tw@}}
\def\obar#1{\overline{\sbox\tw@{$#1$}\dp\tw@\z@\box\tw@}}
\makeatother

\usepackage{bm}
\usepackage{caption}
\usepackage{subcaption}

\makeatletter\let\p@subfigure\thefigure\makeatother

\floatstyle{plaintop}
\restylefloat{table}

\usepackage{fancyhdr} 
\usepackage{cleveref}
\crefname{chapter}{Chapter}{Chapters}
\crefname{section}{Section}{Sections}
\crefname{subsection}{Section}{Sections}
\crefname{subsubsection}{Section}{Sections}
\crefname{figure}{Figure}{Figures}
\crefname{table}{Table}{Tables}
\crefname{equation}{Equation}{Equations}
\crefname{appendix}{Appendix}{Appendices}
\crefname{appendices}{Appendix}{Appendices}

\crefname{appsec}{Appendix}{Appendices}

\usepackage{catoptions}
\makeatletter

\def\Autoref#1{%
  \begingroup
  \edef\reserved@a{\cpttrimspaces{#1}}%
  \ifcsndefTF{r@#1}{%
    \xaftercsname{\expandafter\testreftype\@fourthoffive}
      {r@\reserved@a}.\\{#1}%
  }{%
    \ref{#1}%
  }%
  \endgroup
}
\def\testreftype#1.#2\\#3{%
  \ifcsndefTF{#1autorefname}{%
    \def\reserved@a##1##2\@nil{%
      \uppercase{\def\ref@name{##1}}%
      \csn@edef{#1autorefname}{\ref@name##2}%
      \autoref{#3}%
    }%
    \reserved@a#1\@nil
  }{%
    \autoref{#3}%
  }%
}

\makeatother
\def\equationautorefname~#1\null{%
  Eq.~(#1)\null
}
\def\equationautorefname~#1\null{
Eq.~(#1)\null
}

\usepackage[utf8, latin1]{inputenc}                 

\title{\textbf{Measuring the Effectiveness of US Monetary Policy during the COVID-19 Recession}}
\author[1, 3]{\MakeUppercase{Martin Feldkircher}}
\author[2]{\MakeUppercase{Florian Huber}\thanks{Corresponding author: Florian Huber. Salzburg Centre of European Union Studies, University of Salzburg. \textit{Address}: M\"{o}nchsberg 2a, 5020 Salzburg, Austria. \textit{Email}: \href{mailto:florian.huber@sbg.ac.at}{florian.huber@sbg.ac.at}. Florian Huber and Michael Pfarrhofer gratefully acknowledge financial support from the Austrian Science Fund (FWF, grant no. ZK 35).}}
\author[2]{\MakeUppercase{Michael Pfarrhofer}}   
\affil[1]{\textit{Oesterreichische Nationalbank (OeNB)}}
\affil[2]{\textit{University of Salzburg}}
\affil[3]{\textit{Vienna School of International Studies (DA)}}
\date{\today}
\begin{document}
\maketitle
\thispagestyle{empty}\normalsize\vspace*{-2em}\small
\begin{center}
\begin{minipage}{0.8\textwidth}
\noindent\small \textbf{Abstract}.
The COVID-19 recession that started in March 2020 led to an unprecedented decline in economic activity across the globe. To fight this recession, policy makers in central banks engaged in expansionary monetary policy. This paper asks whether the measures adopted by the US Federal Reserve (Fed) have been effective in boosting real activity and calming financial markets. To measure these effects at high frequencies, we propose a novel mixed frequency vector autoregressive (MF-VAR) model. This model allows us to combine weekly and monthly information within an unified framework. Our model combines a set of macroeconomic aggregates such as industrial production, unemployment rates and inflation with high frequency information from financial markets such as stock prices, interest rate spreads and  weekly information on the Feds balance sheet size. The latter set of high frequency time series is used to dynamically interpolate the monthly time series to obtain weekly macroeconomic measures. We use this setup to simulate counterfactuals in absence of monetary stimulus. The results show that the monetary expansion caused higher output growth and stock market returns, more favorable long-term financing conditions and a depreciation of the US dollar compared to a no-policy benchmark scenario. 
\end{minipage}
\end{center}

\newpage\doublespacing
\section{Introduction}\label{sec:introduction}

Worldwide restrictions to contain the spread of the novel Coronavirus (COVID-19) triggered a sharp drop in global economic activity, a collapse in trade and a severe rise in unemployment. First estimates for 2020 point at considerable contractions of GDP in most advanced economies \citep{McKibbin2020}. Policymakers responded swiftly, with unprecedented fiscal stimulus packages in the magnitude of nearly 15\% of global GDP.\footnote{For an overview of the enacted policy measures, see \url{https://www.bruegel.org/publications/datasets/covid-national-dataset/}} In the same vein, central banks provided stimulus by loosening their policy stance considerably. In many emerging economies, central banks successfully introduced forms of quantitative easing for the first time \citep{Arslan2020,Hartley2020}, while in advanced economies with policy space, easings took mostly the form of rate cuts, which further facilitated the use of fiscal stimulus packages.     

In the US, the economic effect of the pandemic was felt strongly on labor markets: employment dropped sharply and wages were cut \citep{Cajner2020,Kurmann2020}. This weakened demand and inflation considerably. The negative business climate also deterred financial markets, with equity prices collapsing more strongly than in any previous  crises triggered by infectious disease outbreaks \citep{Baker2020}. Relatedly, US Treasury markets experienced a sharp sell-off, leading to spikes in long-term yields \citep{Schrimpf2020}. The US Federal Reserve (Fed) responded with several measures including the opening of credit facilities to support malfunctioning markets and actions aimed at relieving cash-flow stress for small and medium-sized businesses, as well as municipalities. The most prominent actions, however, were moving the policy rate back towards the zero lower bound and resuming the monthly purchase of massive amounts of securities.

This paper tries to give a first assessment of how successful the monetary easing in the US was in stabilizing prices and providing stimulus to the economy. One concern when assessing the effectiveness of policy responses in real-time is the low frequency nature of many macroeconomic aggregates (with most of them available on a monthly or quarterly frequency, at best).  Even if we rely on monthly data we are left with only very few observations that we can use to infer the effects of monetary policy during the COVID-19 crisis on several key quantities of interest for policy makers.

For that purpose, we borrow strength from data which is available at higher frequencies. These time series are often sampled at daily or weekly frequency and allow us to construct weekly measures of industrial production, inflation and unemployment. This is achieved within a coherent multivariate framework that allows for dynamic interactions between the macroeconomic and financial quantities considered.

Our proposed econometric framework is a mixed frequency vector autoregression (MF-VAR) which models all variables on a weekly frequency. Using a state space representation of the multivariate system, we recast the lower frequency quantities in terms of a weekly component with missings between monthly observed values. These missing observations are subsequently estimated by taking into account the properties of the model and using the higher frequency time series dynamically. Our model is then used to simulate the effects of monetary policy shocks. Using these shocks we can compute weekly historical decompositions and perform counterfactual scenarios to investigate the effects the monetary policy measures had on the US economy.

Our results indicate that without a monetary expansion, US economic activity would have been significantly lower. In other words, the US Fed, so far, has been successful in cushioning the economic consequences of the COVID-19 crisis. Positive effects on output growth are underpinned by a rise in stock market returns, an easing of long-term financing conditions and a depreciation of the US dollar. By contrast, effects on inflation and the unemployment rate are statistically insignificant. 
 
The remainder of this paper is structured as follows. Section \ref{sec:econometrics} briefly describes the dataset and econometric model used while Section \ref{sec:results} shows the main results. In this section, we discuss the dynamic reactions to a monetary policy shock and discuss the historical decompositions. Finally, the last section briefly summarizes and concludes the paper.

\section{Empirical framework}\label{sec:econometrics}

\subsection{A Mixed Frequency VAR Model}
As stated in the introductory section, one key issue with adequately assessing the impacts of COVID-19 related monetary policy measures is the extremely short time span of available data. To provide a timely estimate, one could focus on high frequency variables such as interest rate spreads or stock prices. But these are typically not of direct interest for policy makers. In policy making circles, assessing the effects of monetary policy interventions on output, inflation and labor markets is pertinent. Unfortunately, for all these variables we only have a handful of observations, rendering an adequate assessment of policy effectiveness difficult.

As a solution, we propose pairing a panel of weekly indicators, contained in an $M_H$-dimensional vector $\bm y_{t, H}$, with monthly indicators stored in an $M_L$-dimensional vector $\bm y_{t, L}$ in a MF-VAR. These vectors run from $t=1, \dots, T$, with $T$ denoting the number of weeks in our sample. Following \cite{schorfheide2015real}, we assume that $\bm y_{t, H}$ is a latent weekly measure of the low frequency indicator.

One key objective is to infer $\bm y_{H, t}$ to obtain weekly measures of the low frequency variables. This is achieved by defining $\bm y_t = (\bm y_{t, L}, \bm y_{t, H})'$, which is an $M (= M_H + M_L)$-dimensional vector, and assuming that it follows a VAR($P$) process:
\begin{equation}
    \bm y_t = \bm A_1 \bm y_{t-1} + \dots + \bm A_p \bm y_{t-p} + \bm \varepsilon_t, \quad \bm \varepsilon_t \sim \mathcal{N}(\bm 0_M,   \bm \Sigma_t) \label{eq: MFBVAR}
\end{equation}
where $\bm A_j ~(j=1, \dots, M)$ are $M \times M$ coefficient matrices associated with lags $j = 1,\dots, P$. $\bm \varepsilon_t$ is a white noise Gaussian process with variance-covariance matrix $\bm \Sigma_t$ that varies over time. To speed up computation and assume that the Covid-19 shock led to a sharp increase in the conditional variance of all elements in $\bm y_t$, we introduce a common stochastic volatility (CSV) model originally proposed in \cite{carriero2016common}. This implies that $\bm \Sigma_t$ is driven by a scalar factor such that:
\begin{equation*}
    \bm \Sigma_t = e^{h_t} \times \bm \Sigma.
\end{equation*}
We assume that $h_t$ evolves according to an AR(1) process:
\begin{equation*}
    h_t = \mu_h + \rho_h (h_{t-1} - \mu_h) + \sigma_h v_t, \quad v_t \sim \mathcal{N}(0, 1).
\end{equation*}
Here, $\mu_h$ denotes the unconditional mean, $\rho_h$ the autoregressive parameter and $\sigma^2_h$ the error variance. $h_t$  simply scales the time-invariant variance-covariance matrix $\bm \Sigma$. This allows us to capture sudden common shifts in variances while leaving the contemporaneous relations unchanged over time.  

Equation (\ref{eq: MFBVAR}) can be cast in its companion form:
\begin{align}
    \bm z_t = \bm F \bm z_{t-1} + \bm \eta_t, \label{eq: companion}
\end{align}
with $\bm z_t = (\bm y'_t, \dots, \bm y'_{t-P+1})'$ and $\bm F$ being the $K \times K$  companion matrix (for $K=PM$) with the first $M$ rows given by $(\bm A_1, \dots, \bm A_p)$. The remaining rows are defined to return an identity such that $\bm y_{t-j} = \bm y_{t-j}$ for $j = 1, \dots, P-1$.  The first $M$ elements of $\bm \eta_t$ are equal to $\bm \varepsilon_t$ while the remaining elements are equal to zero.

The missing values in $\bm y_t$ can be obtained by interpreting (\ref{eq: companion}) as a state evolution equation that provides information on how the  elements in $\bm z_t$ (and thus $\bm y_t$) are related over time. Following much of the recent literature \citep{koop2020uk, koop2020regional, gefang2020computationally}, we assume that the four-week-average of $\bm y_{L, t}$ which we denote by $\tilde{\bm x}_{L, t}$, is related to $\bm y_{L, t}$ as follows:
\begin{equation*}
 \tilde{\bm x}_{t, L} = \left(\bm y_{L, t} + \bm y_{L, t-1} + \bm y_{L, t-2} +\bm y_{L, t-3}\right)/4.
\end{equation*}
This equation states that we view $\tilde{\bm x}_{t, L}$ as the (observed) average of the weekly latent indicators. Notice that this assumption implies that each month features exactly four weeks (and thus we drop four weeks per year to arrive at $48$ weeks). Define a selection matrix $\bm S_{L t}$ that equals an identity matrix in time $t$ only in the last week of a month while being equal to a zero matrix for the initial three weeks, and $\bm \Lambda_L$ is a matrix such that:
\begin{equation*}
\bm x_{t, L} =  \bm S_{L, t}  \tilde{\bm x}_{t, L} = \bm S_{L, t} \bm \Lambda_L \bm z_t.
\end{equation*}

For the monthly indicators, we assume that the identity $\bm x_{t, H} = \bm y_{t, H}$ holds if the dataset is balanced. If some monthly values are missing, we introduce a separate selection matrix $\bm S_{M, t}$ with $\bm x_{M, t} = \bm S_{M, t} \bm y_{M, t}$.

Following \citet{schorfheide2015real}, the observation equation that relates the observed to the latent quantities is:
\begin{equation}
\bm x_t = \bm M_t  \bm \Lambda  \bm z_t.
\end{equation}
Here, $\bm x_t = (\bm x'_{t, L}, \bm x'_{t, H})'$, $\bm M_t$ is a selection matrix and $\bm \Lambda$ is composed of $\bm \Lambda_L$ and appropriate selection vectors to single out the high frequency quantities in $\bm z_t$.

We estimate the MF-VAR using Bayesian techniques. This implies that we need to specify suitable priors on all parameters of the model. In this paper, we use the conjugate Minnesota prior on the VAR coefficients which has also been used by \cite{schorfheide2015real}. On the remaining model parameters (which comprise of the parameters of the state equation of $h_t$ and $\bm \Sigma$), we use a Beta prior on the autoregressive coefficient $\rho_h$, a normally distributed prior on the unconditional mean $\mu_h$ and a Gamma prior on $\sigma^2_h$. Finally, we use an weakly informative inverse Wishart prior on $\bm \Sigma$. Estimation is carried out using the Markov chain Monte Carlo (MCMC) algorithm discussed in \cite{schorfheide2015real} and efficiently implemented in the \texttt{R} package \textbf{mfbvar}\citep{ankargren2019mixed}.

\subsection{Data}\label{subsec:data}
Our analysis focuses on the reaction of the consumer price index (\texttt{CPIAUCSL}), the unemployment rate (\texttt{UNRATE}) and industrial production (\texttt{INDPRO}) to a monetary policy easing. All of these focal variables are on a monthly frequency.  Higher-frequency variables consist mainly of financial indicators. In particular, we include the money supply (\texttt{M2}) as the policy variable, the \textit{five-year forward inflation expectation rate} (\texttt{T5YIFR}) to gauge market-based inflation expectations, the NASDAQ composite indicator (\texttt{NASDAQCOM}), the \textit{US dollar/euro foreign exchange rate} (\texttt{DEXUSEU}) and the \textit{ten-year treasury constant maturity rate} (\texttt{WGS10YR}). As measures of financial stress we rely on the \textit{CBOE volatility index} (VIX, \texttt{VIXCLS}). 

The sample period runs from the first week of 2011 to week $24$ of 2020 (end of week: June 8, 2020) and is taken from the FRED database of the Federal Reserve Bank of St. Louis (\href{https://fred.stlouisfed.org}{fred.stlouisfed.org}). If the raw data for financial variables is on a higher frequency than weekly (that is, daily for \texttt{T5YIFR}, \texttt{NASDAQCOM}, \texttt{DEXUSEU}, \texttt{VIXCLS}), we take the arithmetic average over the respective weekdays. All variables enter the model as year-on-year differences.

\section{Scenario and counterfactual analysis}\label{sec:results}

In this section we examine the effects of an expansion of the US money supply on output, inflation, the unemployment rate and several financial indicators. In what follows, we proceed in two steps. First, we look at the overall plausibility of our model by examining impulse response functions. For that purpose, we rely on a simple recursive identification scheme with ordering the monthly variables first, followed by \texttt{M2}. Last, we put all other weekly indicators. Note that this simple recursive scheme implies zero restrictions on the low-frequency variables. In particular, in our application the Cholesky decomposition implies that there are no contemporaneous effects of the high-frequency indicators on inflation, output and the unemployment rate, an assumption with which most economist would agree upon. 

\begin{figure}
    \centering
    \includegraphics[width=\textwidth]{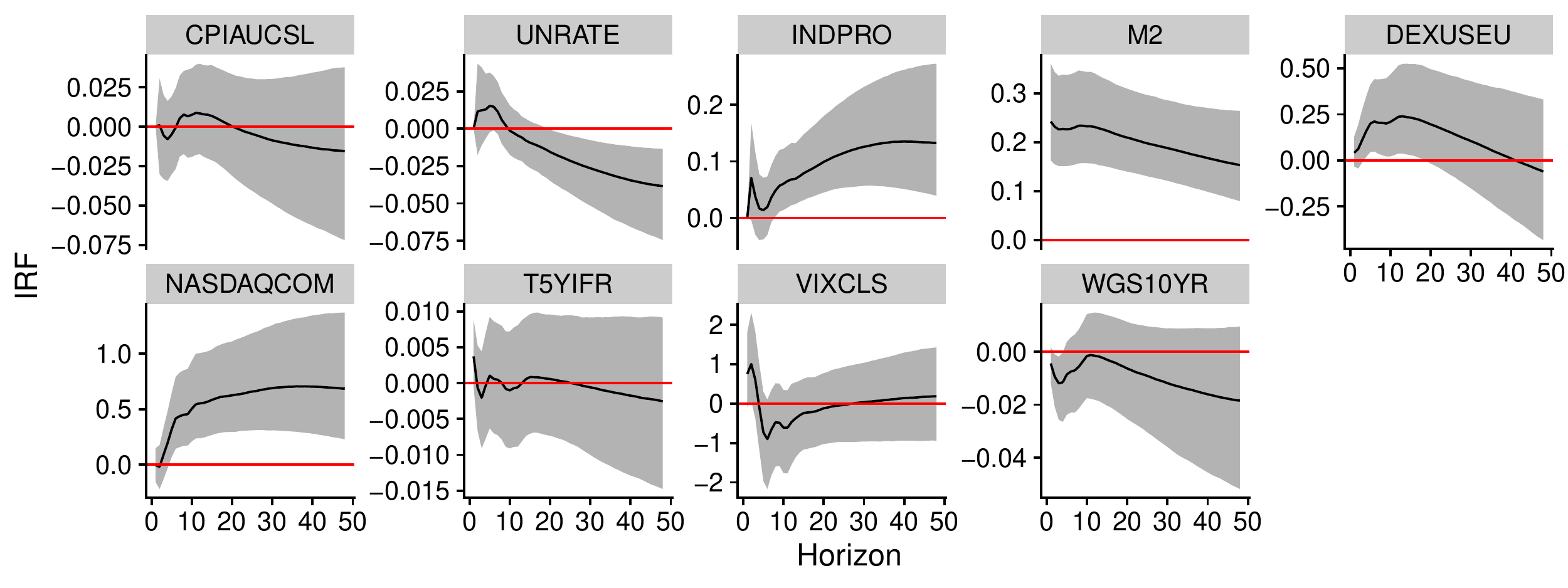}
    \caption{Impulse response functions to a one-standard deviation shock to \texttt{M2}.}\vspace*{-0.5em}
    \caption*{\footnotesize\textit{Notes}: Median response alongside the 90 percent posterior credible set. The red line marks zero.}
    \label{fig:irfs}
\end{figure}

The results are depicted in \autoref{fig:irfs} which shows the posterior median (solid line) along with 90\% credible intervals. The figure demonstrates that the expansionary shock to the money supply 
(\texttt{M2}) significantly drives up output growth and lowers the unemployment rate. These effects are rather persistent and take place with a lag. We do not find a significant upward effect on inflation, although we have included inflation expectations which in general should help mitigating the price puzzle \citep{Castelnuovo2010} often encountered in empirical studies. This finding can be explained by the time period under consideration, which was characterized by low interest and inflation rates. As regards financial variables, we see a a significant and persistent upward effect on equity returns, a front-loaded depreciation of the US dollar and a decrease of long-term yields. Also the VIX increases immediately, which could be related to the positive and pronounced shoot-up of equity returns. Summing up, the mixed-frequency approach generates impulse response functions that are in line with predictions of the bulk of empirical studies on the effects of monetary policy.

Next, we generate counterfactual scenarios. For that purpose, we construct historical decompositions that explain deviations of time series from their trend by shocks to the equations in the system. Neutralizing shocks to money supply after the onset of the COVID-19 crisis thus yields a counterfactual scenario to answer the question how output growth, unemployment and inflation would have evolved without the Fed having provided monetary stimulus. 

\begin{figure}
    \centering
    \includegraphics[width=\textwidth]{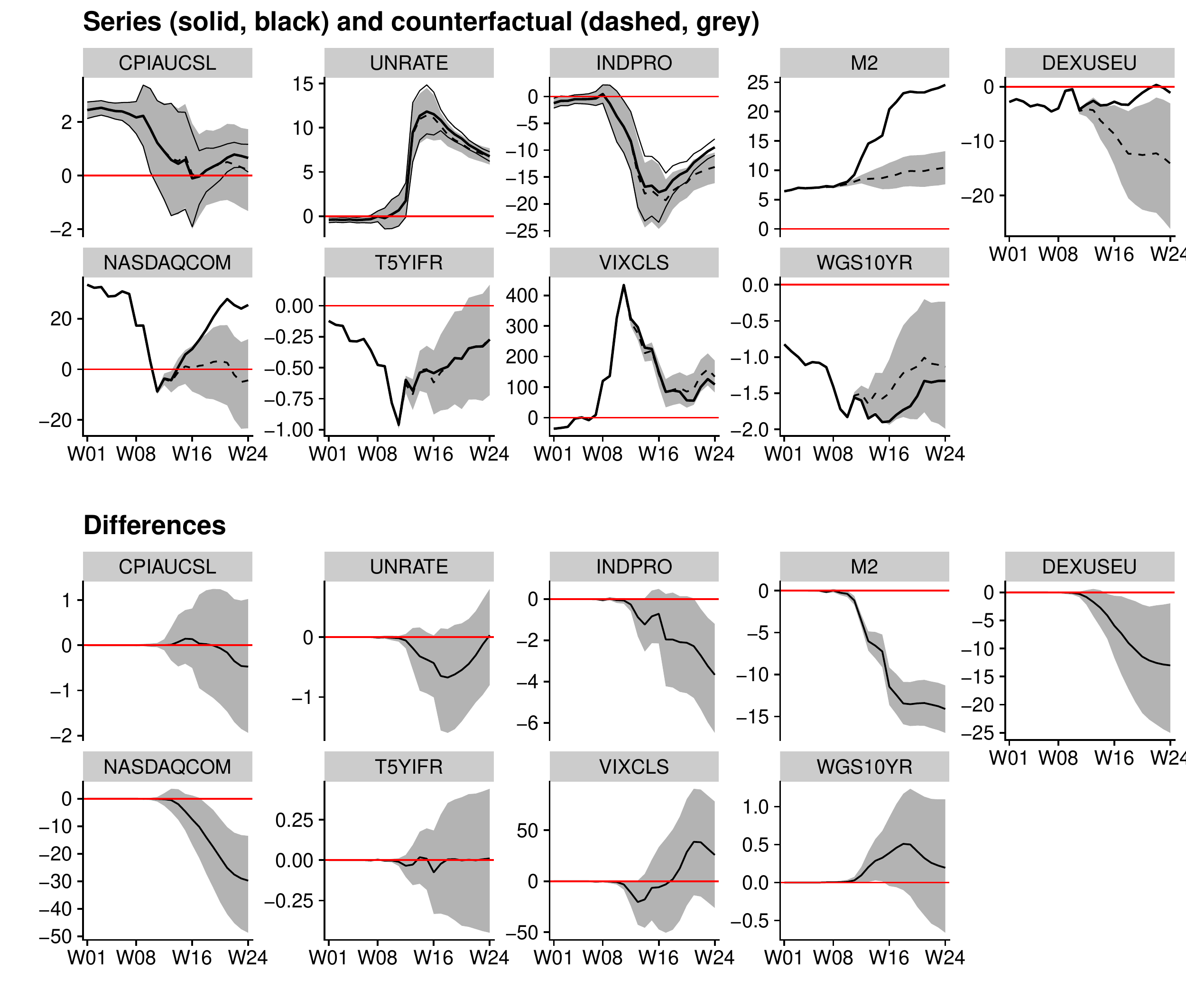}
    \caption{Counterfactual analysis based on setting identified shocks to \texttt{M2} after the onset of the COVID-19 crisis to zero. }
    \vspace*{-0.5em}
    \caption*{\footnotesize\textit{Notes}: Upper panel: The black solid lines depict the actual evolution of the series (alongside the 90 percent posterior credible set for monthly variables), the dashed line alongside the grey shaded area (90 percent posterior credible set) shows the counterfactual. Lower panel: Posterior of the differences between the actual and counterfactual scenario. The red line marks zero.}
    \label{fig:hd}
\end{figure}

The results are depicted in \autoref{fig:hd}. In the upper panels, we show the evolution of actual series (black thick lines) and responses under the counterfactual scenario (grey shaded area, dashed line) along with 90\% credible intervals. Since high-frequency movements of low-frequency variables are estimated within the MF-VAR framework, we also depict credible intervals for the historical weekly evolution of inflation, the unemployment rate and output growth (black thin lines).

The results indicate that output growth would have been weaker without monetary policy stimulus provided by the US Fed. This finding could be driven by the strong effect monetary policy exerted on financial variables: equity returns would have been considerably lower and long-term yields higher under the no-policy scenario. The analysis also suggests that monetary policy triggered a stronger depreciation of the exchange rate and hence a boost to external competitiveness of the US economy. By contrast, the counterfactuals show no significant effect on unemployment and inflation. Considering the delayed response of unemployment discussed in the context of the impulse response functions, this might be an artefact of the considered counterfactual period being to short to detect effects of the expansion yet.

To investigate the significance  more systematically, the bottom panel of \autoref{fig:hd} presents the differences of the responses under the no-policy and the policy scenario along with 90\% credible intervals. 
That analysis corroborates the findings from above that monetary policy led to higher output growth, a pick up in equity returns and an easing in long-term financing conditions. It also led to a significantly lower value of the US dollar.  

\section{Closing remarks}\label{sec:conclusions}
In this note, we gave a first empirical investigation of the effects of US monetary policy to stimulate growth in response to COVID-19. For that purpose, we have estimated a MF-VAR on monthly and weekly data. This model allows us to estimate weekly measures of industrial production, inflation and the unemployment rate. We then simulate the effects of expansionary monetary policy and assess its effects on the endogenous variables in the model.

The results suggest that the US Fed was successful in stimulating growth on the back of higher equity prices and more favorable long-term financing conditions. Also, monetary policy triggered a depreciation of the US dollar supporting external competitiveness of the US economy. By contrast, we do not find significant effects on unemployment and inflation, both variables that typically react more sluggishly to economic stimulus.

\newpage
\small{\setstretch{0.85}
\addcontentsline{toc}{section}{References}
\bibliographystyle{custom.bst}
\bibliography{lit}}

\end{document}